\documentclass[a4paper,10pt,twoside]{cpc-hepnp}
\usepackage{CJK,upgreek,fancyhdr}
\usepackage{multicol}
\usepackage{graphicx}
\usepackage{booktabs}
\usepackage{amssymb,bm,mathrsfs,bbm,amscd}
\usepackage[tbtags]{amsmath}
\usepackage{lastpage}

\usepackage{color}
\usepackage[normalem]{ulem}

\begin{document}

\fancyhead[c]{\small Chinese Physics C~~~Vol. xx, No. xx (201x) xxxxxx}
\fancyfoot[C]{\small 065104-\thepage}

\footnotetext[0]{Received 28 November 201x}

\title{Probing the anisotropic distribution of baryon matter in the Universe using fast radio bursts\thanks{Supported by the National Natural Science Foundation of China (11603005, 11775038, 12005184). }}

\author{%
Hai-Nan Lin$^{1}$
\quad Yu Sang$^{2;1)}$\email{sangyu@yzu.edu.cn (Corresponding author)}
}
\maketitle

\address{%
$^1$Department of Physics, Chongqing University, Chongqing 401331, China\\
$^2$Center for Gravitation and Cosmology, College of Physical Science and Technology, Yangzhou University, Yangzhou 225009, China\\
}

\begin{abstract}
We propose that fast radio bursts (FRBs) can be used as the probes to constrain the possible anisotropic distribution of baryon matter in the Universe. Monte Carlo simulations show that, 400 (800) FRBs are enough to detect the anisotropy at 95\% (99\%) confidence level, if the dipole amplitude is at the order of magnitude 0.01. However, much more FRBs are required to tightly constrain the dipole direction. Even 1000 FRBs are far from enough to constrain the dipole direction within angular uncertainty $\Delta\theta<40^{\circ}$ at 95\% confidence level. The uncertainty on the dispersion measure of host galaxy does not significantly affect the results. If the dipole amplitude is in the level of 0.001, however, 1000 FRBs are not enough to correctly detect the anisotropic signal.
\end{abstract}

\begin{keyword}
radio continuum: transients, cosmological parameters, large-scale structure of Universe
\end{keyword}


\footnotetext[0]{\hspace*{-3mm}\raisebox{0.3ex}{$\scriptstyle\copyright$}2017
Chinese Physical Society and the Institute of High Energy Physics
of the Chinese Academy of Sciences and the Institute
of Modern Physics of the Chinese Academy of Sciences and IOP Publishing Ltd}%

\begin{multicols}{2}

\section{Introduction}\label{sec:introduction}

Fast radio bursts (FRBs) are short-duration and energetic radio transients with typical radiation frequency $\sim$ GHz and typical duration $\sim$ milliseconds happening in the Universe. For recent reviews, see e.g. Ref. \cite{Petroff:2019tty,Zhang:2020qgp,Xiao:2021omr}. The first discovery of FRBs can be traced back to 2007, when Lorimer et al. reanalyzed the archive data of the Parkes 64-m telescope and found an extraordinary radio pulse, which is now named FRB 010724 \cite{Lorimer:2007qn}. This phenomenon has for long time not attracted much attention from astronomers, until four other bursts were discovered in 2013 \cite{Thornton:2013iua}. Since then, FRBs have aroused great interests within the astronomy community. The observed dispersion measures (DM) of most FRBs greatly excess the contribution from the Milky Way, indicating that they occur at cosmological distance. The cosmological origin is further confirmed as the identification of the host galaxy and the direct measurement of redshift \cite{Keane:2016yyk,Chatterjee:2017dqg,Tendulkar:2017vuq}. Thanks to the progress in the observational technique, more and more FRBs have been discovered in recent years. Up to now, hundreds of FRBs have been reported \cite{Petroff:2016tcr,CHIMEFRB:2021srp}. Generally, FRBs can be divided into two types according to whether they are repeating or not. Most FRBs observed so far is apparently non-repeating, and tens of FRBs are found to be repeating but without periodicity. Except for an extremely active repeating source, FRB 121102, from which hundreds of bursts have been detected \cite{Spitler:2016dmz,Scholz:2017kwy,Zhang:2018jux,Gourdji:2019lht}, most of the rest repeating sources found by the Canadian Hydrogen Intensity Mapping Experiment (CHIME) telescope only repeat two or three times \cite{Andersen:2019yex}. Statistical analysis of FRB 121102 shows that the burst energies from this source follow the bent power-law and are scale-invariant \cite{Lin:2019ldn}, implying that there are some similarities between FRBs and soft gamma repeaters (SGRs) \cite{Chang:2017bnb}. Recently, the CHIME/FRB Collaboration \cite{Amiri:2020gno} found an unexpected long period of 16.35 days with an approximately 4-day active window in FRB 180916.J0158+65. Interestingly, the recently discovered FRB 200428 is found to be associated with the Galactic magnetar SGR 1935+2154 \cite{Andersen:2020hvz}, which implies that magnetar can be the progenitor of at least some FRBs.

FRBs are very energetic and detectable up to high redshift \cite{Zhang:2018csb}. Therefore, they can be used as the probes to study the cosmology.  Yu \& Wang showed that FRBs can be used to measure the cosmic proper distance \cite{Yu:2017beg}. Walters et al. showed that FRBs can be used to constrain the cosmological parameters, especially the baryon matter density \cite{Walters:2017afr}.  Li et al. proposed that FRBs can be used to constrain the fraction of baryon mass in the intergalactic medium (IGM) model-independently \cite{Li:2019klc,Li:2020qei}.  Xu \& Zhang proposed that FRBs can be used to probe the intergalactic turbulence \cite{Xu:2020jic}. Wu et al. pointed out that FRBs can be used to measure the Hubble parameter $H(z)$ independent of cosmological model \cite{Wu:2020jmx}. Pagano \& Fronenberg showed that the highly dispersed FRBs can be used to constraining the epoch of cosmic reionization \cite{Pagano:2021zla}. Qiang et al. showed that FRBs can be used test the possible cosmic anisotropy \cite{Qiang:2019zrs}. In addition, the strongly lensed FRBs can be used to probe the compact dark matter in the Universe \cite{Munoz:2016tmg}. Especially, strongly lensed repeating FRBs can tightly constrain the Hubble constant and cosmic curvature \cite{Li:2017mek}. Interestingly, strongly lensed repeating FRBs can be used as the probes to search for gravitational waves \cite{Pearson:2020wxb}. Besides, FRBs can be used to test the fundamental physics, such as constraining the weak equivalent principle and photon mass \cite{Wei:2015hwd,Wu:2016brq,Tingay:2016tgf,Bonetti:2016cpo}.

The observations on the cosmic microwave background (CMB) shows that the Universe is homogeneous and isotropic in large scale \cite{Ade:2015hxq,Akrami:2019bkn}. However, there is no direct evidence that the baryon matter is also isotropic. Especially, observations on the luminosity of type-Ia supernovae show that there is possible anisotropy \cite{Antoniou:2010gw,Bengaly:2015dza,Lin:2016jqp}. In addition, observations on the fine-structure constant from quasar absorbtion line also show that the Universe may be anisotropy \cite{Webb:2010hc,King:2012id}. Interestingly, the dipole fitting to both the supernovae data and fine-structure constant data leads to a consistent dipole direction \cite{Mariano:2012wx}. The anisotropy can be caused by e.g. the interaction of photons with the anisotropic distribution of baryon matter. In this paper, we will show that FRBs can be used to test the possible anisotropic distribution of baryon matter in the Universe. The DM is the integral of electron density alone the line of sight, while the latter is proportional to the baryon matter in the Universe. By measuring the DM of hundreds of FRBs at different directions on the sky, we can detect the anisotropic signal in the baryon matter distribution.

The rest parts of this paper are arranged as follows: In Section 2, we present the methodology of how to using FRBs to test the anisotropic distribution of baryon matter. In Section 3, we investigate the capability of future FRB data in testing the anisotropic distribution of baryon matters using Monte Carlo simulations. Finally, discussions and conclusions are given in Section 4.

\section{Methodology}\label{sec:methodology}

Due to the interaction of photons with free electrons, photons with different energies travel with different speed. The relative time delay between low- and high- energy photons propagating from a distant source to earth is proportional to the dispersion measure (DM), which is the integral of electron density along the photon path \cite{Inoue:2003ga}. This effect is especially important in the low-energy wave bands, such as the radio bands in which FRBs are observed. The DM is related to the matter distribution along the light path, so it contains the information of the Universe and the distance of FRB source.

In general, the observed DM of a FRB contains three parts: the contributions from Milky Way (MW), intergalactic medium (IGM) and host galaxy
\cite{Deng:2013aga,Gao:2014iva},
\begin{equation}
  {\rm DM_{obs}}={\rm DM_{MW}}+{\rm DM_{IGM}}+\frac{{\rm DM_{host}}}{1+z},
\end{equation}
where the factor $1+z$ accounts for the cosmic dilation. The term ${\rm DM_{MW}}$ can be well constrained by modelling the electron distribution of the MW \cite{Taylor:1993my,Cordes:2002wz,Yao_2017msh}, as long as the position of FRB is known. Thus it can be subtracted from the total observed DM, leaving behind the extragalactic DM,
\begin{equation}\label{eq:DM_E_obs}
  {\rm DM_{E}}\equiv{\rm DM_{obs}}-{\rm DM_{MW}}.
\end{equation}
The uncertainty of ${\rm DM_E}$ is propagated from the uncertainties of ${\rm DM_{obs}}$ and ${\rm DM_{MW}}$, i.e. $\sigma_E=\sqrt{\sigma^2_{\rm obs}+\sigma^2_{\rm MW}}$. The ${\rm DM_{obs}}$ can be tightly constrained by observing the time-resolved spectra of FRBs. According to the FRB catalog \cite{Petroff:2016tcr}, the average uncertainty on ${\rm DM_{obs}}$ is only 1.5 pc cm$^{-3}$. For FRBs at high Galactic latitude $(|b|>10^\circ)$, the DM contributing from MW can be tightly constrained with an average uncertainty of 10 pc cm$^{-3}$ \cite{Manchester:2004bp}. Therefore, we take $\sigma_{\rm obs}=1.5 {\rm pc~cm}^{-3}$ and $\sigma_{\rm MW}=10 {\rm pc~cm}^{-3}$ in the following calculations. We treat ${\rm DM_E}$ in equation (\ref{eq:DM_E_obs}) as the observed quantity. On the other hand, if we can model ${\rm DM_{IGM}}$ and ${\rm DM_{host}}$, the extragalactic DM can be also calculated theoretically by
\begin{equation}\label{eq:DM_E_th}
  {\rm DM_{E}^{\rm th}}={\rm DM_{IGM}}+\frac{{\rm DM_{host}}}{1+z}.
\end{equation}
By comparing the observed and theoretical ${\rm DM_E}$, the cosmological parameters can be constrained.

The DM contributing from IGM, assuming that both hydrogen and helium are fully ionized\footnote{This is justified at $z\lesssim 3$ \cite{Meiksin:2007rz,Becker:2010cu}.}, can be written as \cite{Deng:2013aga,Zhang:2020ass}
\begin{equation}\label{eq:DM_IGM}
  {\rm \overline{DM}_{IGM}}(z)=\frac{21cH_0\Omega_b f_{\rm IGM}}{64\pi Gm_p}\int_0^z\frac{1+z}{\sqrt{\Omega_m(1+z)^3+\Omega_\Lambda}}dz,
\end{equation}
where $m_p=1.673\times 10^{-27}$ kg is the proton mass, $f_{\rm IGM}$ is the fraction of baryon in the IGM, $H_0$ is the Hubble constant, $G$ is the gravitational constant, $\Omega_b$ is the normalized baryon matter density, $\Omega_m$ and $\Omega_\Lambda$ are the normalized densities of matter (includes baryon matter and dark matter) and dark energy at present day, respectively. Note that equation (\ref{eq:DM_IGM}) is based on the assumptions that the hydrogen and helium are fully ionized, and the matter fluctuation is negligible. We introduce an uncertainty term $\sigma_{\rm IGM}$ to account for the possible deviation of the actual ${\rm DM_{IGM}}$ from the theoretical expectation. Following Ref. \cite{Yang:2016zbm}, we assume $\sigma_{\rm IGM}=100 {\rm pc~cm}^{-3}$. In order to test the possible anisotropic distribution of baryon matter in the Universe, we model the baryon density as the dipole form, i.e., the baryon density at direction $\hat{\mathbf p}$ is given by
\begin{equation}\label{eq:omega_b}
  \Omega_b(\hat{\mathbf p})=\Omega_{b0}(1+A{\hat{\mathbf n}}\cdot{\hat{\mathbf p}}),
\end{equation}
where $\Omega_{b0}$ is the mean baryon density, $A$ is the dipole amplitude, $\hat{\mathbf{n}}$ is the dipole direction which can be parameterized by the longitude and latitude $(l,b)$ in the galactic coordinates. In this case, ${\rm DM_{IGM}}$ not only depends on the redshift $z$, but also depends on the direction ${\hat{\mathbf p}}$ of FRB source in the sky, i.e.,
\begin{eqnarray}\label{eq:DM_IGM_aniso}\nonumber
  {\rm \overline{DM}_{IGM}}({\hat{\mathbf p}},z)=\frac{21cH_0\Omega_{b0}f_{\rm IGM}}{64\pi Gm_p}(1+A{\hat{\mathbf n}}\cdot{\hat{\mathbf p}}) \\ \times\int_0^z\frac{1+z}{\sqrt{\Omega_m(1+z)^3+\Omega_\Lambda}}dz,
\end{eqnarray}

Due to the lack of FRBs with identified host galaxy, the local environment of FRB source is still poorly known. Hence it is difficult to model the DM contributing from the host galaxy. Many factors can affect ${\rm DM_{host}}$, such as the type of galaxy, the departure of FRB source from galaxy center, the inclination of host galaxy, etc. Here we model ${\rm DM_{host}}$ according to the evolution of star formation rate (SFR) history \cite{Luo:2018tiy} ,
\begin{equation}\label{eq:DM_host}
  {\rm \overline{DM}_{host}}(z)={\rm DM_{host,0}}\sqrt{\frac{{\rm SFR}(z)}{{\rm SFR}(0)}},
\end{equation}
where ${\rm DM_{host,0}}$ and ${\rm SFR}(0)$ are the DM of host galaxy and SFR at present day, respectively. The SFR evolves with redshift as \cite{Yuksel:2008cu}
\begin{equation}\label{eq:SFR}
  {\rm SFR}(z)=0.02\left[(1+z)^{a\eta}+\left(\frac{1+z}{B}\right)^{b\eta}+\left(\frac{1+z}{C}\right)^{c\eta}\right]^{1/\eta},
\end{equation}
where $a=3.4$, $b=-0.3$, $c=-3.5$, $B=5000$, $C=9$ and $\eta=-10$. We follow Ref. \cite{Li:2019klc} and take ${\rm DM_{host,0}}$ as a free parameter. Equation (\ref{eq:DM_host}) should be also interpreted as the mean contribution from host galaxy. We introduce an uncertainty term $\sigma_{\rm host}$ to account for the possible deviation from the mean value. Following Ref. \cite{Yang:2016zbm}, we consider two different fiducial values for ${\rm DM_{host,0}}$ and its uncertainty, i.e. $({\rm DM_{host,0}},\sigma_{\rm host})=(100,20)~{\rm pc~cm^{-3}}$ and $({\rm DM_{host,0}},\sigma_{\rm host})=(200,50)~{\rm pc~cm^{-3}}$, respectively.

By fitting the observed ${\rm DM_E}$ to the theoretical ${\rm DM_E}$, the cosmological parameters can constrained. The best-fitting parameters are the ones which can minimize the $\chi^2$,
\begin{equation}
  \chi^2=\sum_{i=1}^N\left[\frac{({\rm DM_E}-{\rm DM_E^{th}})^2}{\sigma_{\rm total}^2}\right],
\end{equation}
where the observed ${\rm DM_E}$ is given by equation (\ref{eq:DM_E_obs}), the theoretical ${\rm DM_E^{th}}$ is given by equation (\ref{eq:DM_E_th}), and the total uncertainty is given by \cite{Li:2019klc}
\begin{equation}\label{eq:sigma_total}
  \sigma_{\rm total}=\sqrt{\sigma_{\rm obs}^2+\sigma_{\rm MW}^2+\sigma_{\rm IGM}^2+\sigma_{\rm host}^2/(1+z)^2}.
\end{equation}
Note that the Hubble constant $H_0$, the mean baryon density $\Omega_{b0}$, and the fraction of baryon mass $f_{\rm IGM}$ are completely degenerated with each other. These three parameters appear together as their product, as is seen in equation (\ref{eq:DM_IGM_aniso}). Therefore, we take the product $h_0\Omega_{b0}f_{\rm IGM}$ as a free parameter, where $h_0\equiv H_0/(100~{\rm km~s^{-1}~Mpc^{-1}})$. In addition, $\Omega_m$ depicts the background Universe and it has been tightly constrained by Planck data, hence we fix it to the Planck 2018 value \cite{Aghanim:2018eyx}. This finally leaves five free parameters $(h_0\Omega_{b0}f_{\rm IGM},A,l,b,{\rm DM_{host,0}})$.

\section{Monte Carlo simulations}\label{sec:simulations}

In this section, we use the Monte Carlo simulations to investigate the ability of future FRB observations in constraining the possible anisotropic distribution of baryon matter. We work in the fiducial $\Lambda$CDM cosmology with the Planck 2018 parameters \cite{Aghanim:2018eyx}, i.e. $H_0=67.4~{\rm km~s^{-1}~Mpc^{-1}}$, $\Omega_m=0.315$, $\Omega_\Lambda=0.685$ and $\Omega_{b0}=0.0493$. The fraction of baryon in the IGM is taken to be $f_{\rm IGM}=0.84$ \cite{Li:2020qei}. Considering the anisotropy of baryon matter, we take a fiducial dipole amplitude $A=0.01$, and without loss of generality, the dipole direction is arbitrary chosen to be $(l,b)=(180^{\circ},0^{\circ})$.

Due to the poor knowledge on physical mechanism and the lack of direct redshift measurement, the actual redshift distribution of FRB is still unclear. Yu \& Wang assumed that the redshift distribution of FRBs is similar to that of gamma-ray bursts \cite{Yu:2017beg} . Li et al. assumed that FRBs have a constant comoving number density, but with an exponential cutoff \cite{Li:2019klc}. Here we assume that the intrinsic event rate density of FRBs follows the SFR, where the redshift distribution of FRBs takes the form \cite{Zhang:2020ass}
\begin{equation}\label{eq:pdf_redshift}
  P(z)\propto\frac{4\pi D^2_c(z){\rm SFR}(z)}{(1+z)H(z)},
\end{equation}
where $D_c(z)=\int_0^zc/H(z)dz$ is the comoving distance, $H(z)=H_0\sqrt{\Omega_m(1+z)^3+\Omega_\Lambda}$ is the Hubble expansion rate, and ${\rm SFR}(z)$ is given by equation (\ref{eq:SFR}). As for the sky direction, since most FRBs are extragalactic origin, they are expected to be uniformly distributed in the sky.

We simulate $N$ FRBs, each contains the following parameters: the redshift $z$, the direction of FRB in the galactic coordinates $(l',b')$, the extragalactic DM and the total uncertainty $({\rm DM_{E}},\sigma_{\rm total})$. The detailed procedures of simulation are as follows:
\begin{enumerate}
  \item The redshift $z$ is randomly sampled according the probability density function given in equation (\ref{eq:pdf_redshift}). The upper limit of redshift is set to $z_{\rm max}=3$ in the simulation.
  \item The sky direction $(l',b')$ is randomly sampled from the uniform distribution, i.e., $l'\sim {\mathcal U}(0^{\circ},360^{\circ})$ and $b'\sim {\mathcal U}(-90^{\circ},90^{\circ})$.
  \item Calculate the fiducial and anisotropic ${\rm \overline{DM}_{IGM}}$ according to equation (\ref{eq:DM_IGM_aniso}). Then randomly sample ${\rm DM_{IGM}}$ from the Gaussian distribution, ${\rm DM_{IGM}}\sim {\mathcal N}({\rm \overline{DM}_{IGM}},\sigma_{\rm IGM})$.
  \item Calculate the fiducial values ${\rm \overline{DM}_{host}}$ according to equation (\ref{eq:DM_host}), and randomly sample ${\rm DM_{host}}$ from the Gaussian distribution, ${\rm DM_{host}}\sim {\mathcal N}({\rm \overline{DM}_{host}},\sigma_{\rm host})$.
  \item Calculate the extragalactic DM according to equation (\ref{eq:DM_E_th}), and calculate the total uncertainty $\sigma_{\rm total}$ according to equation (\ref{eq:sigma_total}).
\end{enumerate}

We simulate $N=100,200,300,...,1000$ FRBs respectively, and then use the simulated data points to constrain the free parameters $(h_0\Omega_{b0}f_{\rm IGM},A,l,b,{\rm DM_{host,0}})$. Figure \ref{fig:contour} shows the posterior probability density functions and the 2-dimensional marginalized contours of the free parameters in one simulation for $N=800$, where the fiducial values of ${\rm DM_{host,0}}$ and $\sigma_{\rm host}$ are 100 and 20 ${\rm pc~cm^{-3}}$, respectively. It is shown that the parameters can be tightly constrained and the best-fitting values are consistent with the fiducial values within $1\sigma$ uncertainty.

\end{multicols}
\begin{center}
 \includegraphics[width=0.85\textwidth]{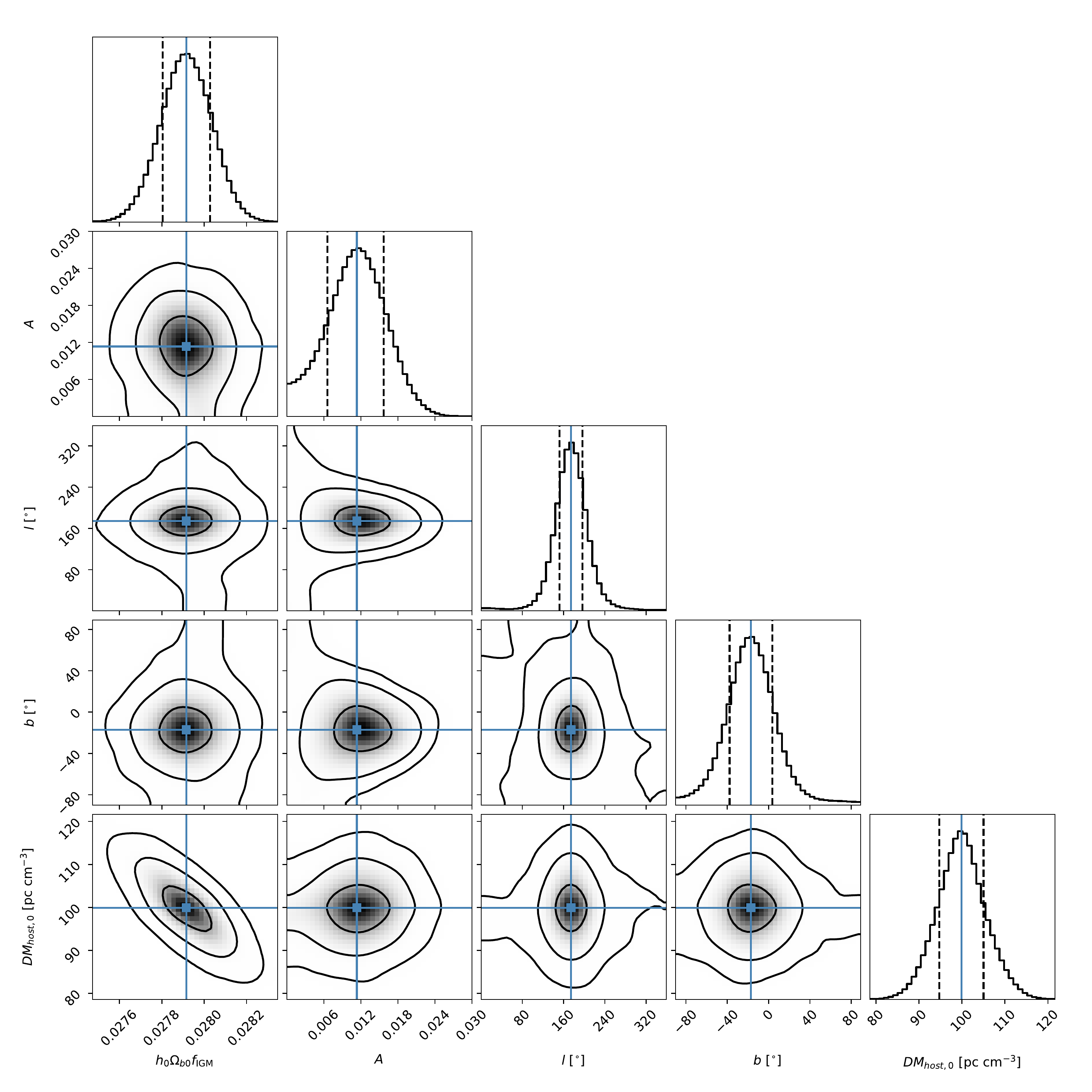}
\figcaption{\label{fig:contour}The best-fitting results in a typical simulation with $N=800$. The fiducial parameters are $A=0.01$, ${\rm DM_{host,0}=100pc~cm^{-3}}$ and $\sigma_{\rm host,0}=20{\rm pc~cm^{-3}}$. The blue solid and black dashed lines stand for the mean value and the $1\sigma$ uncertainty of the parameters, respectively.}
\end{center}
\begin{multicols}{2}

Due to the statistical fluctuation, the best-fitting parameters differ in each simulation. Therefore, we simulate 1000 times for each $N$ with different random seeds. Figure \ref{fig:parameters} shows the results for $N=800$. The upper-left panel shows the best-fitting dipole amplitudes and $1\sigma$ uncertainties in 1000 simulations. The red solid, red dashed and red dotted lines represent the mean value, the fiducial value and the zero value of dipole amplitudes, respectively. The grey and blue error bars mean that the dipole amplitudes are inconsistent and consistent with zero, respectively. We see that only 8 out of the 1000 dipole amplitudes are consistent with zeros within $1\sigma$, hence we say that there is 99.2\% probability that we can detect the anisotropic signal, i.e. $P(A>0)=99.2\%$. The upper-right panel shows the histogram of the dipole amplitudes, which can be well fitted by the Gaussian distribution, with mean value $\bar{A}=(1.14\pm0.01)\times 10^{-2}$ and standard deviation $\sigma_{A}=(0.36\pm0.01)\times 10^{-2}$. The lower-left panel shows the best-fitting dipole directions (black dots) in 1000 simulations. The red plus centering at $(180^{\circ},0^{\circ})$ is the fiducial direction. The two black circles represent the circular regions of radius $\Delta\theta < 20^{\circ}$ and $\Delta\theta<40^{\circ}$ with respect to the fiducial direction, which encircle 3.0\% and 11.7\% of the whole sky, respectively. We find that 372 and 834 best-fitting directions fall into the areas $\Delta\theta<20^{\circ}$ and $\Delta\theta<40^{\circ}$, respectively, i.e. $P(\Delta\theta<20^{\circ})=37.2\%$ and $P(\Delta\theta<40^{\circ})=83.4\%$. The lower-right panel shows the best-fitting $(h_0\Omega_{b0}f_{\rm IGM},{\rm DM}_{\rm host,0})$ values, which are highly correlated.

\end{multicols}
\begin{center}
 \includegraphics[width=0.45\textwidth]{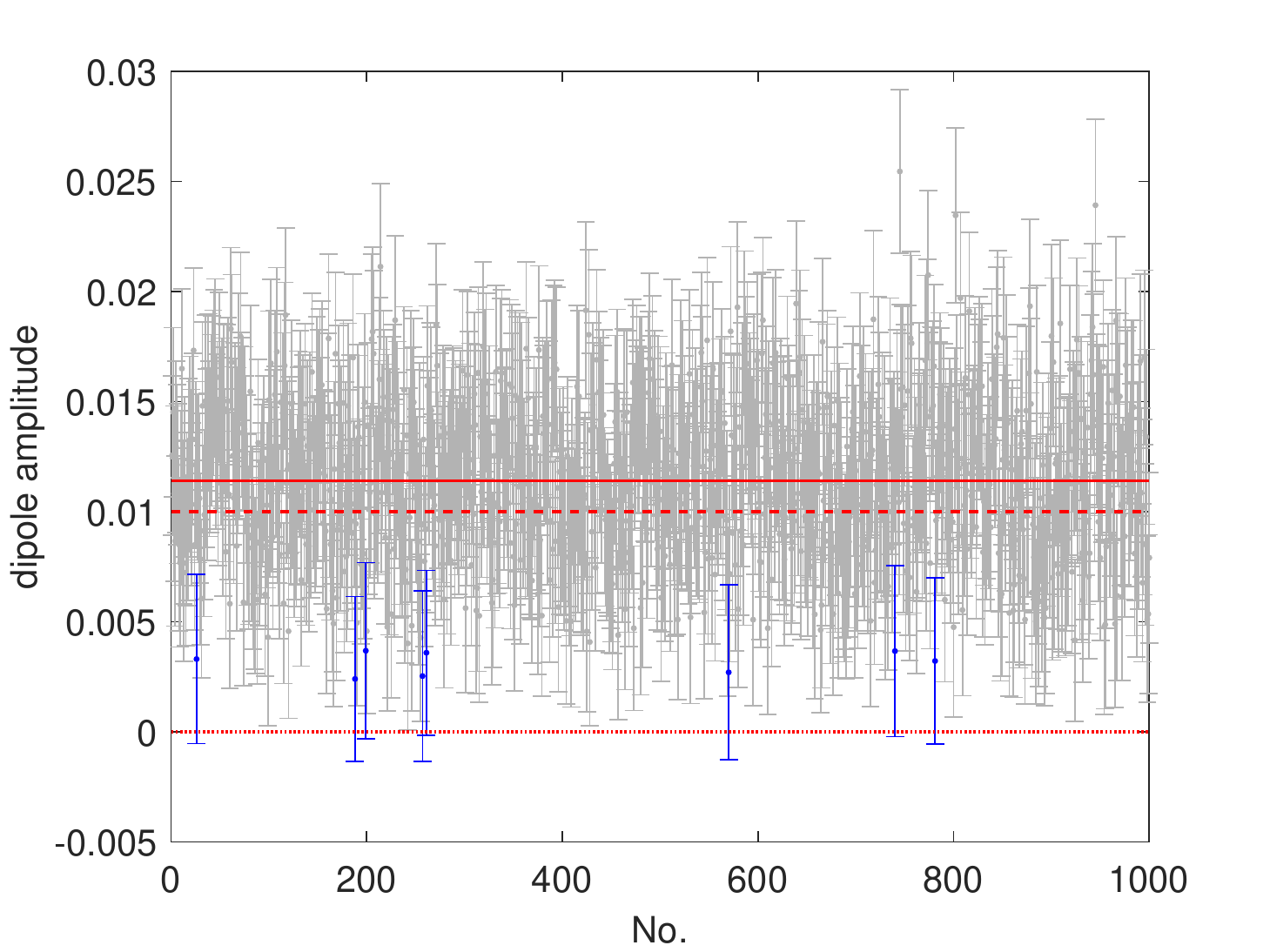}
 \includegraphics[width=0.45\textwidth]{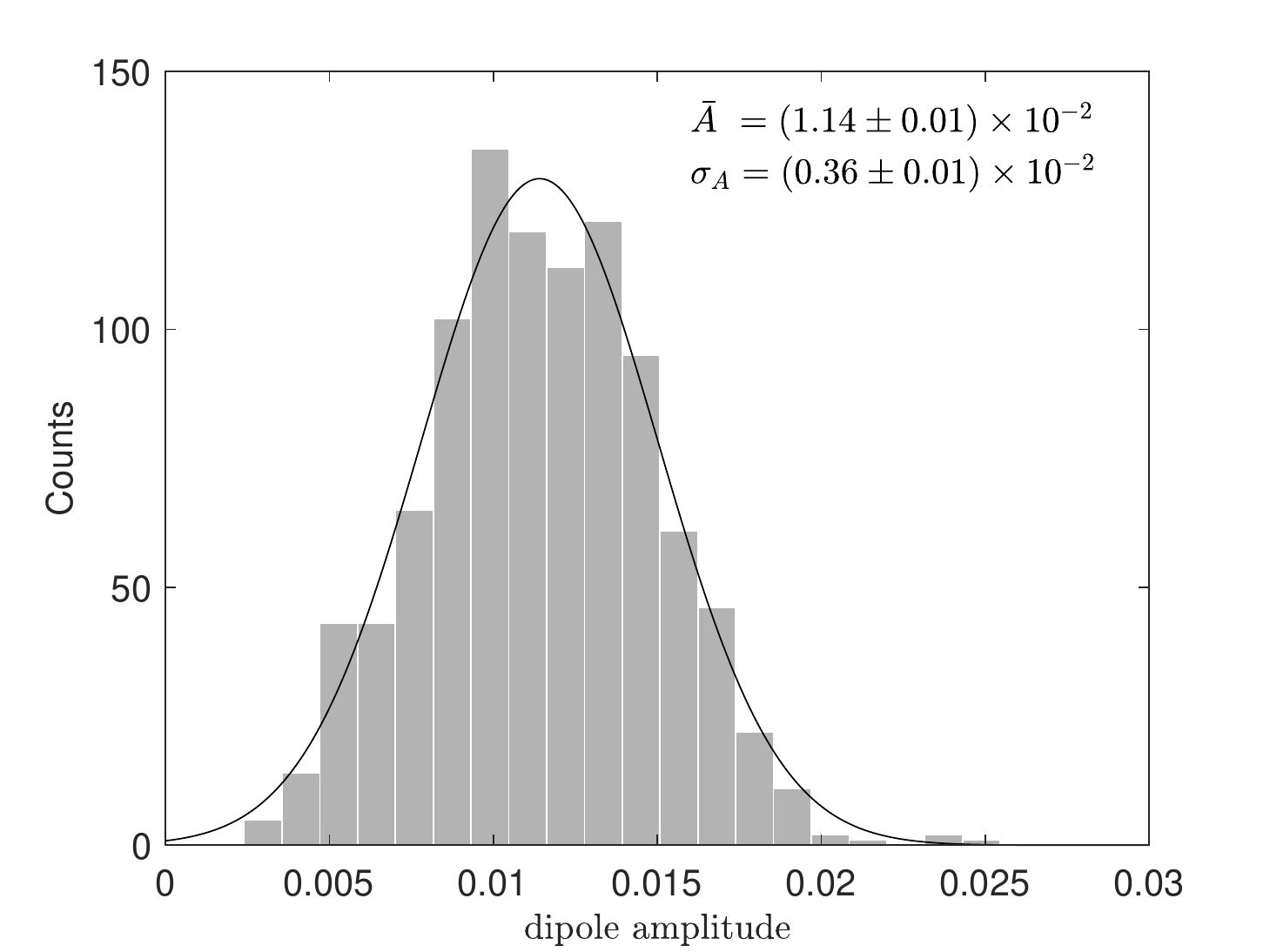}
 \includegraphics[width=0.45\textwidth]{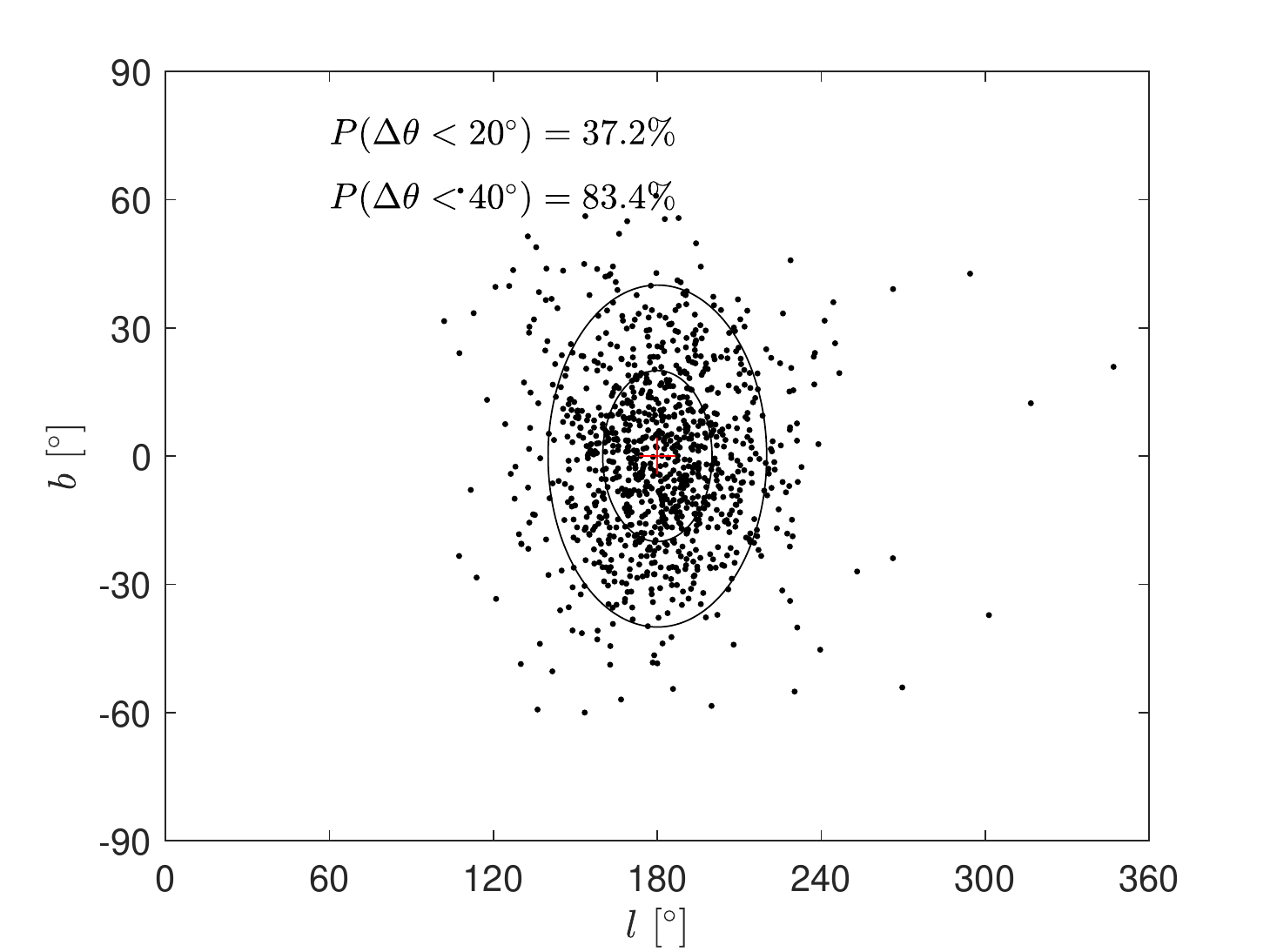}
 \includegraphics[width=0.45\textwidth]{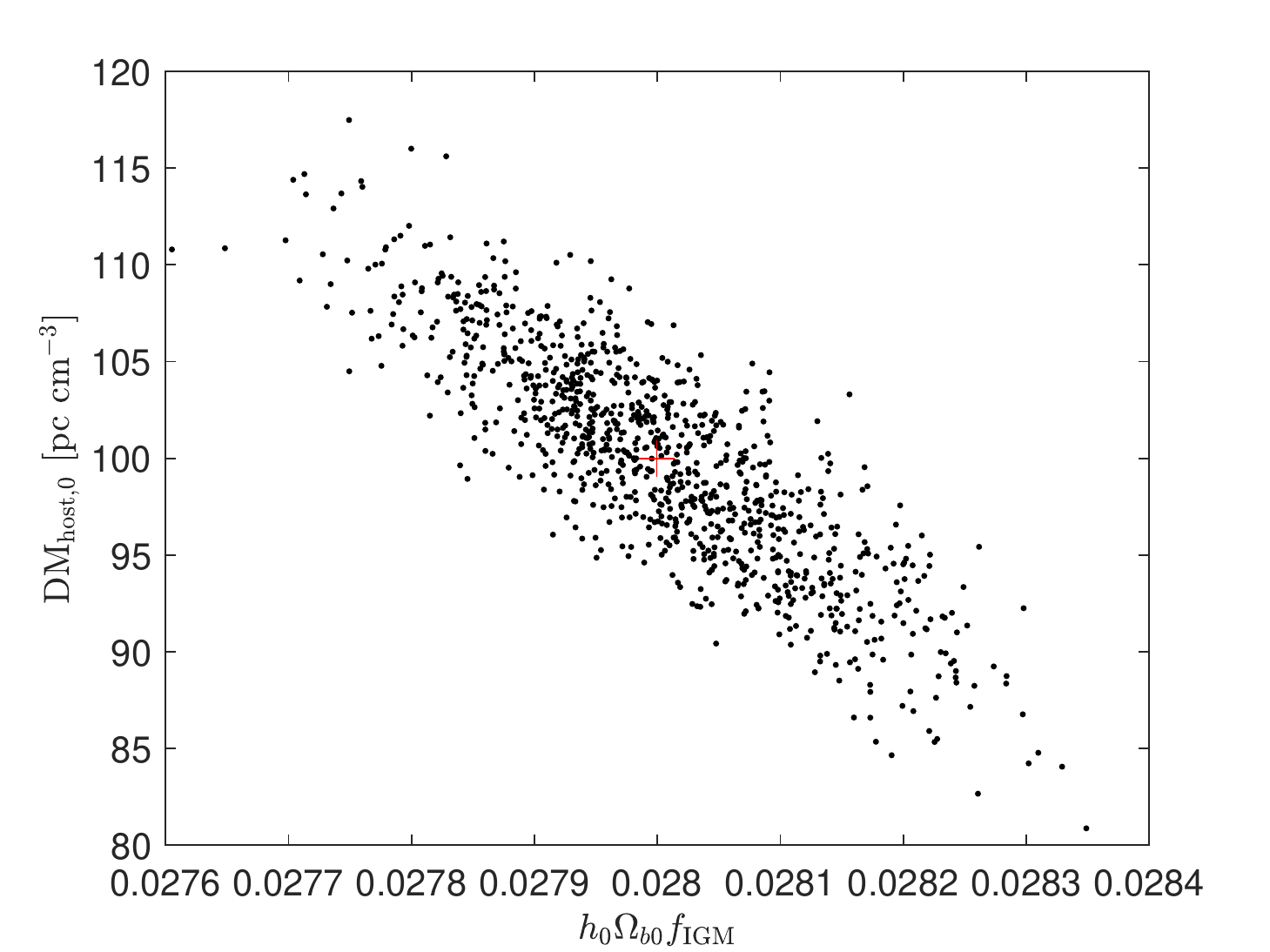}
 \figcaption{The best-fitting parameters in 1000 simulations with $N=800$. The fiducial parameters are $A=0.01$, ${\rm DM_{host,0}=100pc~cm^{-3}}$ and $\sigma_{\rm host,0}=20{\rm pc~cm^{-3}}$. Upper-left: the best-fitting dipole amplitudes $A$ with their $1\sigma$ uncertainty. The horizontal axis is the serial number of simulation.} The red solid, red dashed and red dotted lines represent the mean value, the fiducial value and the zero value of dipole amplitudes, respectively. The grey and blue error bars mean that the dipole amplitudes are inconsistent and consistent with zero, respectively. Upper-right: the histogram of the best-fitting amplitudes. The black line is the best-fitting result to Gaussian distribution. Lower-left: the best-fitting dipole directions $(l,b)$. The red plus centering at $(180^{\circ},0^{\circ})$ is the fiducial direction. The two black circles represent the circular regions of radius $\Delta\theta < 20^{\circ}$ and $\Delta\theta<40^{\circ}$ with respect to the fiducial direction, respectively. Lower-right: the best-fitting $(h_0\Omega_{b0}f_{\rm IGM},{\rm DM}_{\rm host,0})$ values. The red plus is the fiducial value.
 \label{fig:parameters}
\end{center}
\begin{multicols}{2}

We do similar calculation for different values of $N$, and list the results in Table \ref{tab:parameters1}. From this table, we can see that as $N$ increases, the probability that we can detect the anisotropy, i.e. $P(A>0)$, also increases. At the same time, the probabilities $P(\Delta\theta<20^{\circ})$ and $P(\Delta\theta<40^{\circ})$ also increase. This tendency can be seen more clearly from the right panel of Figure \ref{fig:probability}, where we plot the probabilities as a function of $N$. For $N\geq 200$, we can detect the anisotropic signal at more than 90\% confidence level. With about 400 (800) FRBs, the dipole amplitude can be constrained at 95\% (99\%) confidence level. However, in order to correctly found the dipole direction, much more FRBs are required. Even for $N=1000$, $P(\Delta\theta<20^{\circ})$ is no more than 50\%. In order to constrain the dipole direction within $40^{\circ}$ uncertainty at 80\% confidence level, about 700 to 800 FRBs are required. From Table \ref{tab:parameters1}, we can also see that as $N$ increases, the mean value of dipole amplitude $\bar{A}$ goes closer to the fiducial value. For $N\geq 200$, the mean dipole amplitude $\bar{A}$ is consistent with the fiducial value within $1\sigma$ uncertainty, see also the left panel of Figure \ref{fig:probability}, where we plot $\bar{A}$ as a function of $N$.

\end{multicols}
\begin{center}
  \centering
  \tabcaption{The results of 1000 simulations for different values of $N$. The fiducial parameters are $A=0.01$, ${\rm DM_{host,0}=100pc~cm^{-3}}$ and $\sigma_{\rm host,0}=20{\rm pc~cm^{-3}}$. First column: the number of FRBs in each simulation. Second column: the probability that we can detect a non-zeros dipole amplitude. Third and fourth columns: the probabilities that best-fitting dipole direction is consistent with the fiducial direction within $20^{\circ}$ and $40^{\circ}$, respectively. Fifth and sixth columns: the mean value and standard deviation of the dipole amplitudes, respectively.}
  \label{tab:parameters1}
  \begin{tabular*}{100mm}{cccccc}
  \hline\hline
  $N$	& $P(A>0)$ &	$P(\theta<20^{\circ})$ &	$P(\theta<40^{\circ})$	& $A/10^{-2}$	& $\sigma_A/10^{-2}$\\
  \hline
  100&	0.862&	0.093&	0.329&	1.86&	0.86\\
  200&	0.904&	0.141&	0.449&	1.51&	0.67\\
  300&	0.933&	0.198&	0.567&	1.34&	0.53\\
  400&	0.953&	0.234&	0.638&	1.25&	0.50\\
  500&	0.968&	0.272&	0.685&	1.19&	0.44\\
  600&	0.979&	0.293&	0.744&	1.17&	0.41\\
  700&	0.985&	0.355&	0.794&	1.17&	0.39\\
  800&	0.992&	0.372&	0.834&	1.14&	0.36\\
  900&	0.996&	0.388&	0.854&	1.12&	0.34\\
  1000&	0.991&	0.427&	0.864&	1.07&	0.31\\
  \hline
  \end{tabular*}
\end{center}
\begin{multicols}{2}

\end{multicols}
\begin{center}
 \centering
 \includegraphics[width=0.45\textwidth]{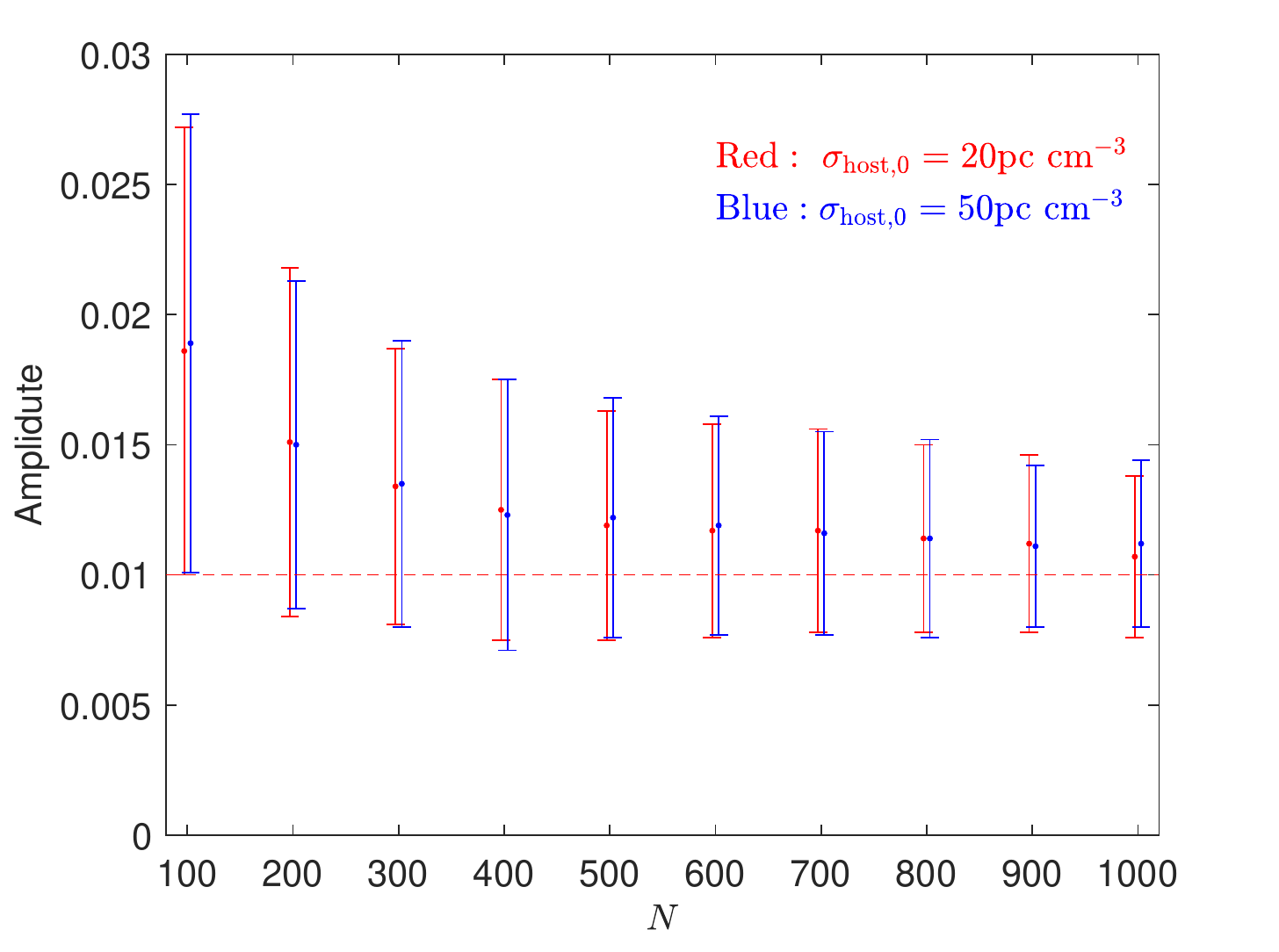}
 \includegraphics[width=0.45\textwidth]{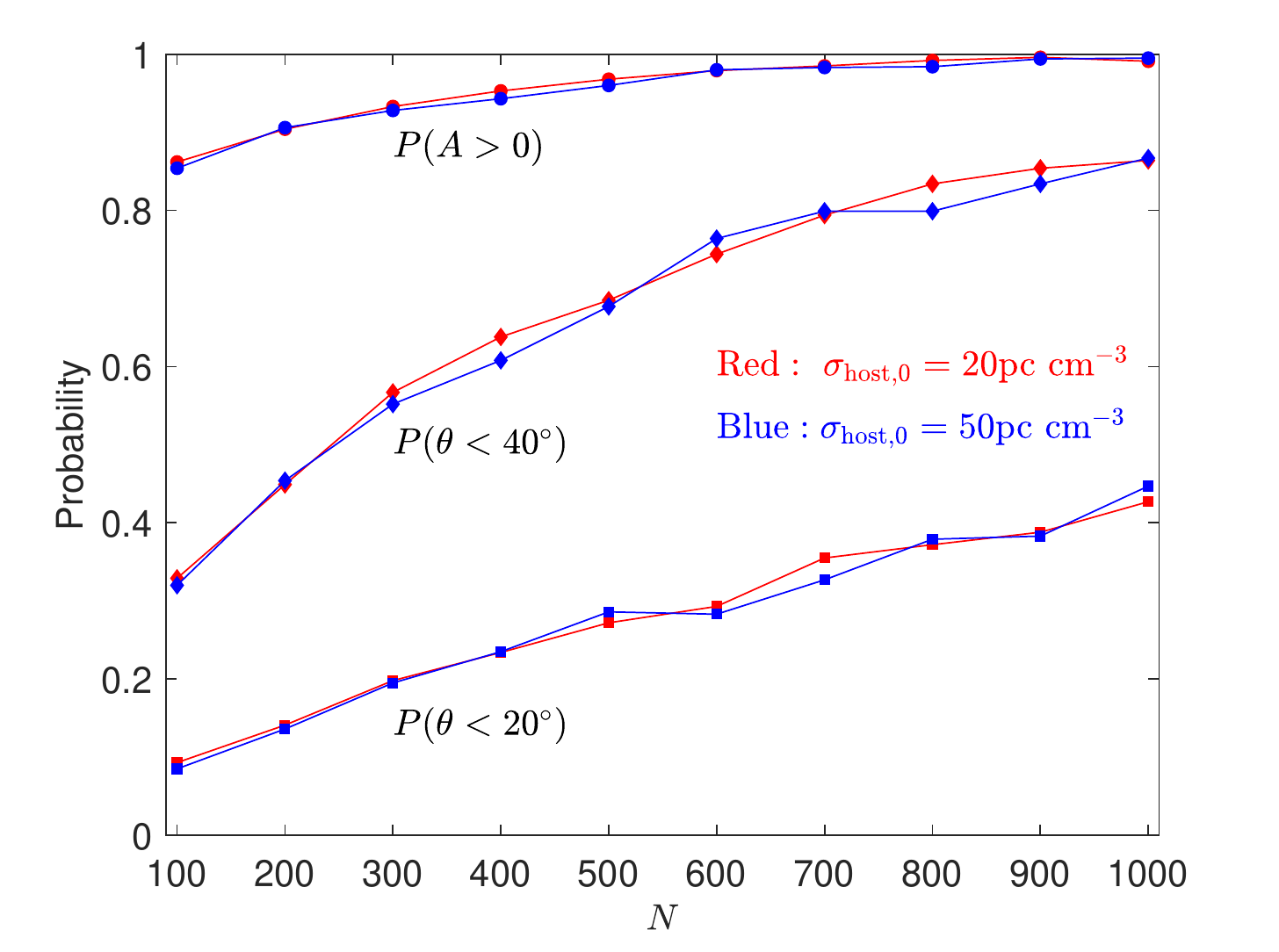}
 \figcaption{Left: the mean value of dipole amplitudes in 1000 simulations as a function of $N$. The error bar represents the standard deviation of dipole amplitudes. Right: the probability that we can correctly reproduce the fiducial dipole amplitude or dipole direction as a function of $N$.}
 \label{fig:probability}
\end{center}
\begin{multicols}{2}

To investigate if the increasing of $\sigma_{\rm host}$ will affect our results or not, we set the fiducial values $({\rm DM_{host,0}},\sigma_{\rm host})=(200,50)~{\rm pc~cm^{-3}}$, and do similar calculation. The results are compared with the case $({\rm DM_{host,0}},\sigma_{\rm host})=(100,20)~{\rm pc~cm^{-3}}$ in Figure \ref{fig:probability}. The left panel plots the mean value of dipole amplitudes in 1000 simulations as a function of $N$, and the right panel shows the probability that we can correctly reproduce the fiducial dipole amplitude or dipole direction as a function of $N$. We see that the increase of $\sigma_{\rm host}$ from 20~pc~cm$^{-3}$ to 50~pc~cm$^{-3}$ almost does not change the results. This can be understood from equation (\ref{eq:sigma_total}): the uncertainty on ${\rm DM_{host}}$ is suppressed by a factor of $1/(1+z)$ due to cosmic dilation, and the total uncertainty is dominated by $\sigma_{\rm IGM}$, which we choose to be 100 pc cm$^{-3}$ in the simulation.

To test if FRBs can probe much weak anisotropic signal, we take a fiducial dipole amplitude $A=0.001$ and carry out similar calculation as above. Figure \ref{fig:parameters2} shows the results of 1000 simulations, with $N=1000$ FRBs in each simulation. The upper-left panel shows the best-fitting dipole amplitudes. We found $P(A>0)=79.2\%$, meaning that there is still $\sim 80\%$ probability that we can detect the anisotropic signal. However, the histogram of dipole amplitudes in the upper-right panel shows that the mean value and standard deviation of the dipole amplitudes is $\bar{A}=(5.14\pm 0.10)\times 10^{-3}$ and $\sigma_A=(2.40\pm 0.10)\times 10^{-3}$, respectively. The mean dipole amplitude is about five times larger than the fiducial dipole amplitude, and it is inconsistent with the fiducial dipole amplitude within $1\sigma$. Especially, the lower-left panel shows that the best-fitting dipole directions are randomly distributed in the sky, $P(\theta<20^{\circ})=4.8\%$ and $P(\theta<40^{\circ})=14.6\%$. This means that we actually can't correctly constrain the dipole anisotropy with 1000 FRBs, if the dipole amplitude is in the level of 0.001.

\end{multicols}
\begin{center}
 \centering
 \includegraphics[width=0.45\textwidth]{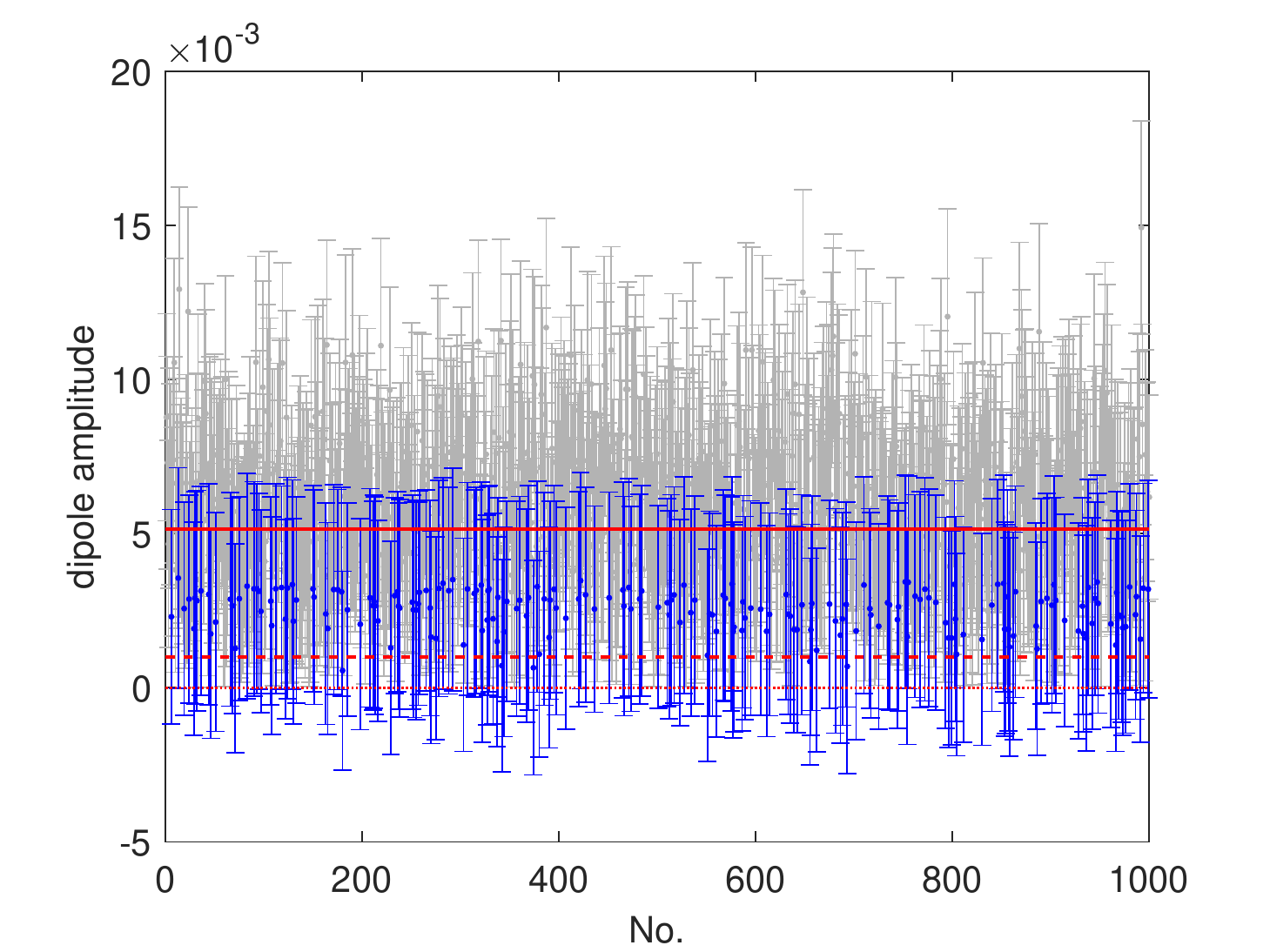}
 \includegraphics[width=0.45\textwidth]{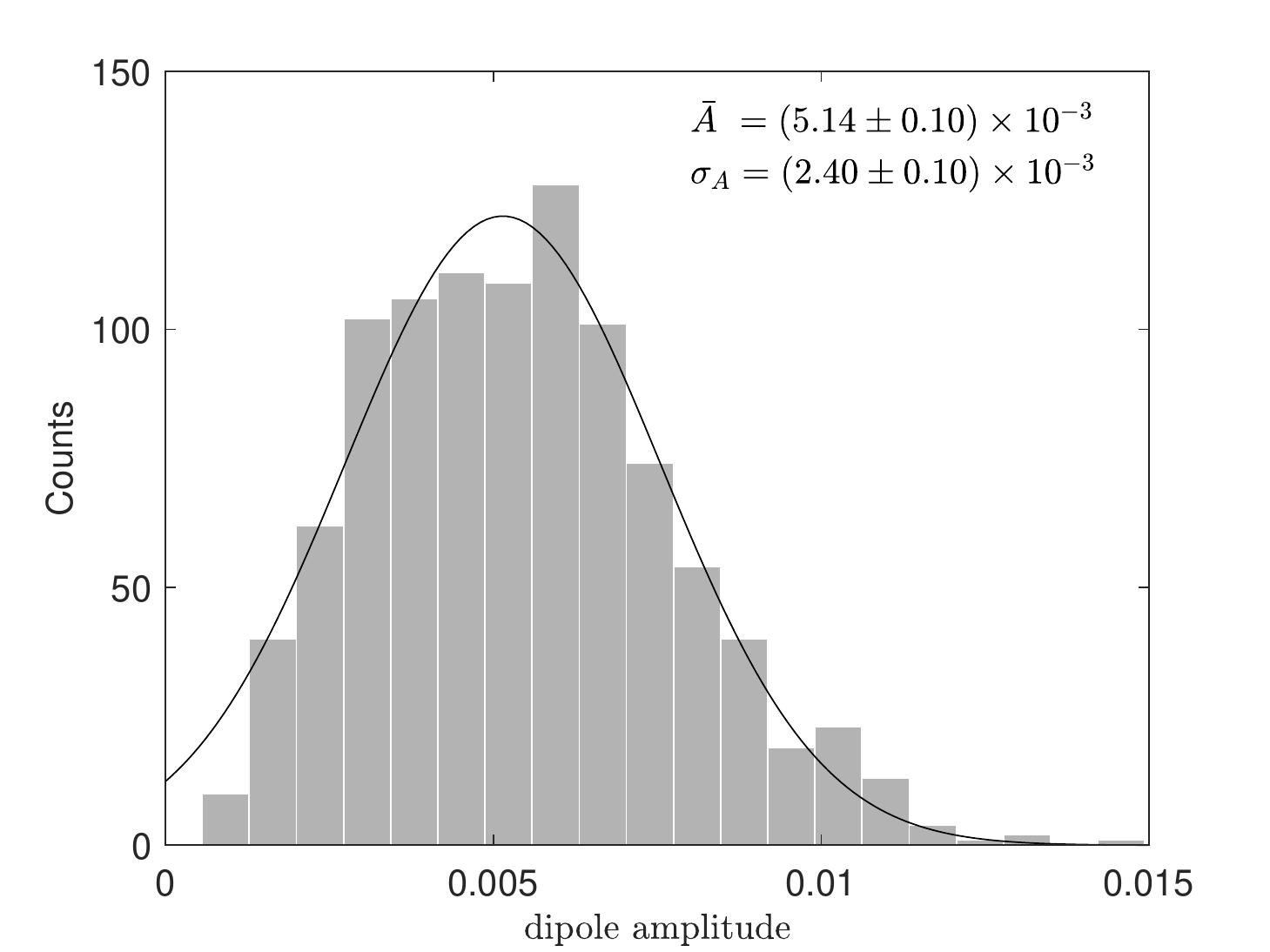}
 \includegraphics[width=0.45\textwidth]{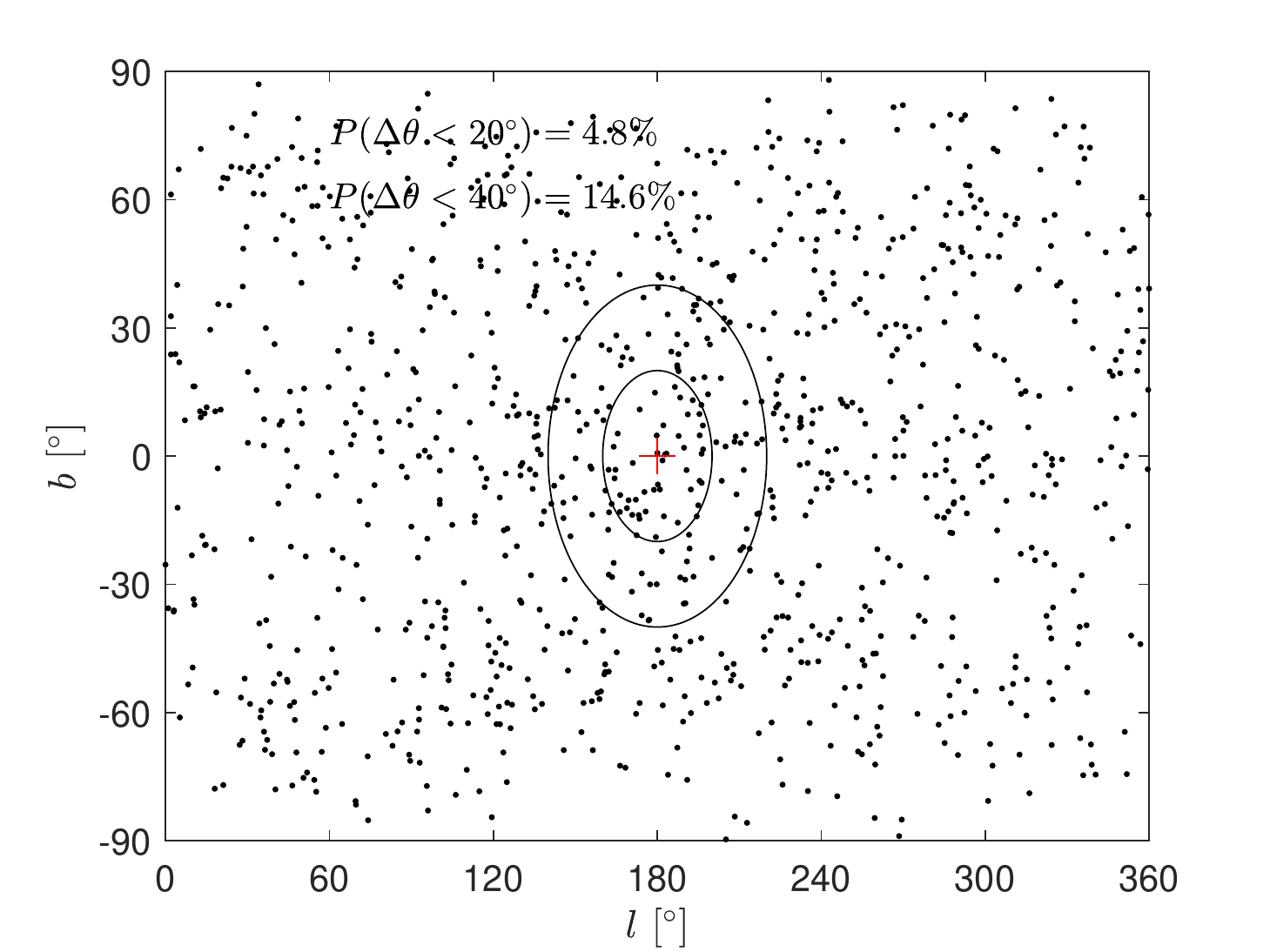}
 \includegraphics[width=0.45\textwidth]{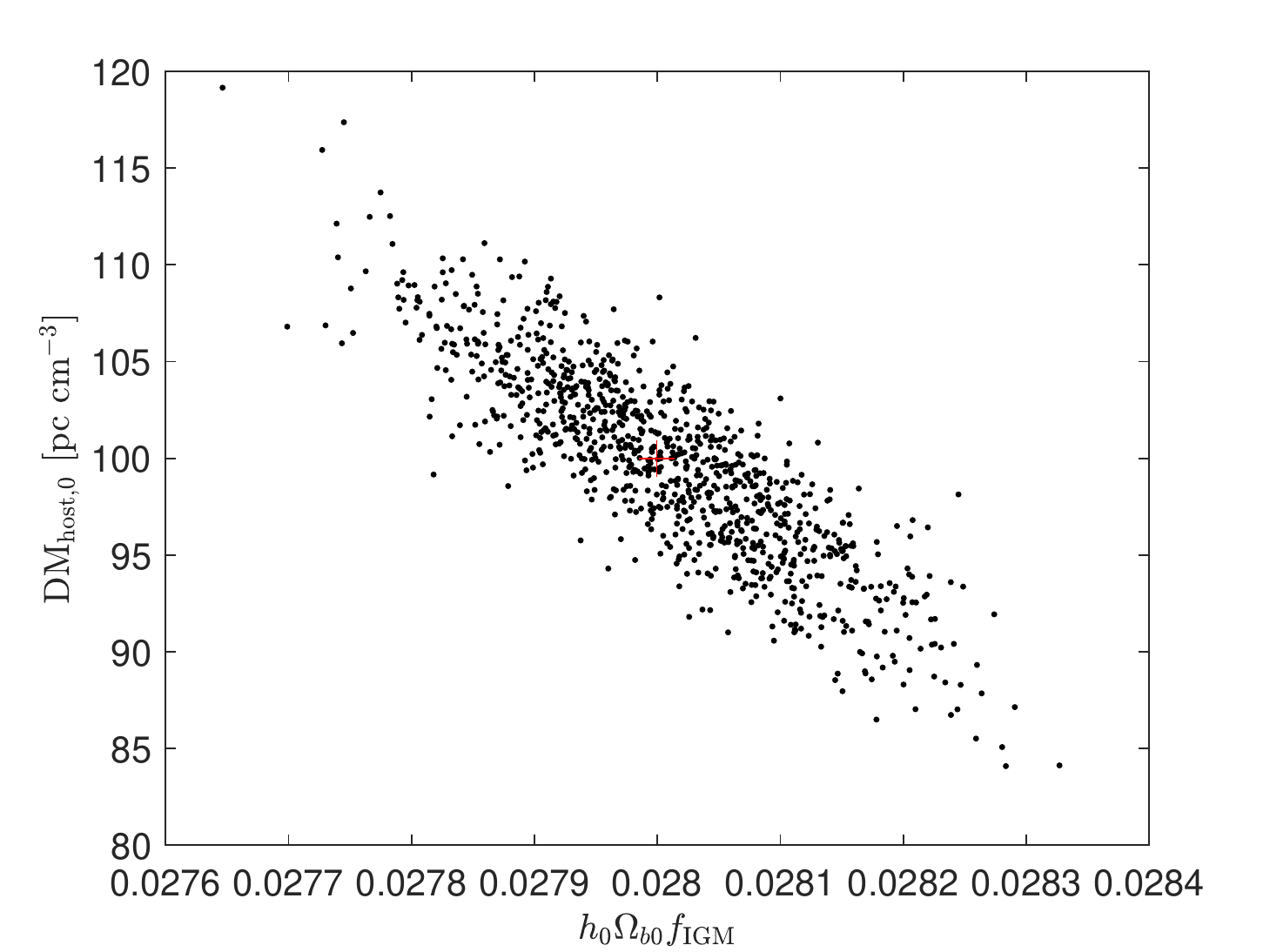}
 \figcaption{The same to Figure \ref{fig:parameters}, but with $N=1000$ and the fiducial dipole amplitude $A=0.001$.}
 \label{fig:parameters2}
\end{center}
\begin{multicols}{2}

\section{Discussions and Conclusions}\label{sec:conclusions}

In this paper, we have investigated the use of FRBs in probing the possible anisotropic distribution of baryon matter in the Universe. We assumed that the distribution of baryon matter has the dipole form, and the fiducial dipole amplitude is chosen to be 0.01. According the simulations, 200 FRBs with well measured redshift and localization are enough to tightly constrain the anisotropy amplitude at 90\% confidence level. With 800 FRBs, the dipole amplitude can be constrained at 99\% confidence level. However, much more FRBs are required to constrain the dipole direction. To constrain the dipole direction with uncertainty $\Delta\theta<40^{\circ}$ (which covering 11.7\% of the whole sky) at 80\% confidence level, we require 700 to 800 FRBs, but even 1000 FRBs are not enough to improve the confidence level to 90\%. These results do not strongly affected by the uncertainty on the DM of host galaxy. If the fiducial dipole amplitude is 0.001, however, 1000 FRBs are far from enough to correctly detect the anisotropic signal.

In a recent paper, Qiang et al. have used the FRBs to test the anisotropy of the Universe use the similar simulation method, but found a very different result \cite{Qiang:2019zrs}. They found that about $N=2800$ FRBs are required to find the cosmic dipole with amplitude 0.01, comparing with $N=200$ FRBs required based on our calculation. The difference may be caused by some reasons. First, we use a different redshift distribution in the simulations. Qiang et al. assumed an exponential distribution $P(z)\propto z\exp(-z)$ similar to that of gamma-ray bursts, while we assumed that the redshift distribution follows the SFR. Compared with the exponential distribution, the redshift distribution we used here has more FRBs at high redshift. Second, Qiang et al. directly assumed that ${\rm DM_E}$ takes the dipole form, while we assumed that the baryon matter density $\Omega_b$ takes the dipole form. Actually, the dipole of $\Omega_b$ is equivalent to the dipole of ${\rm DM_{IGM}}$, since the latter is proportional to the former. But it is not equivalent to the dipole of ${\rm DM_E}$, because ${\rm DM_{\rm host}}$ is redshift-, and may be direction-dependent. Third, Qiang et al. used a six-parameter fit, which is one more parameter (the matter density $\Omega_m$) than we used here. We fix $\Omega_m$ to the Planck value, which is equivalent to adopt the Planck prior. Finally, Qiang et al. have not considered the statistical fluctuations. Their conclusion was based on one simulation for a given $N$. In some case, a much smaller number of FRBs can still find the anisotropic signal.

The progenitors of FRBs are still unclear. The most popular models evolve one or two compact objects (such as neutron stars and magnetars) in the central of FRB source \cite{Petroff:2019tty}. FRBs are expected to be frequent events in the Universe, although some of them can't be observed due to dim luminosity. Based on the compact binary merger model, the event rate of FRBs is estimated to be about $(3\--6)\times 10^4~{\rm Gpc}^{-3}~{\rm yr}^{-1}$ above the energy threshold $E_{\rm th}=3\times 10^{39}$ erg \cite{Cao:2018yzp}. With the running of new radio telescopes, such as the Australian Square Kilometre Array Pathfinder (ASKAP) \cite{Shannon:2018mdj}, the Five-hundred-meter Aperture Spherical Telescope (FAST) \cite{Nan:2011um}, the Canadian Hydrogen Intensity Mapping Experiment (CHIME ) \cite{Amiri:2018qsq}, the BAO from Integrated Neutral Gas Observations (BINGO) \cite{Abdalla:2021yfg}, etc., more FRBs with well measured redshift can be observed. We expected that the anisotropic signal of baryon matter can be detected or ruled out in the near future.

\acknowledgments{This work has been supported by the National Natural Science Foundation of China under Grant Nos. 11603005, 11775038 and 12005184.}

\end{multicols}
\vspace{2mm}
\centerline{\rule{80mm}{0.1pt}}
\vspace{2mm}
\begin{multicols}{2}

\end{multicols}

\clearpage
\end{document}